\documentclass[fleqn,twoside]{article}
\usepackage{espcrc2}
\usepackage[varg]{txfonts}
\usepackage{graphicx}

\newcommand{\g}{\gamma}
\newcommand{\ksea}{\kappa_{\mathrm{sea}}}
\newcommand{\kval}{\kappa_{\mathrm{val}}}
\newcommand{\as}{\alpha_s}
\newcommand{\bk}{B_K}
\newcommand{\bkh}{\hat{B}_K}
\newcommand{\sbar}{\overline{s}}
\newcommand{\msbar}{\overline{\mathrm{MS}}}
\newcommand\gev{\,\mathrm{Ge\kern-0.1em V}}
\renewcommand{\vec}[1]{\mathbf{#1}}
\newcommand{\vp}{\vec p}

\title{Sea quark effects in $\bk$ from $N_f=2$ clover-improved Wilson
fermions}

\author{UKQCD Collaboration:
        J.M.~Flynn\address[SU]{School of Physics \& Astronomy,
          University of Southampton, SO17 1BJ, UK}\thanks{Presenter},
        F.~Mescia\address{Dipartimento di Fisica, Univ. degli Studi
          ``Roma TRE", via della Vasca Navale 84, I-00146 Roma,
          Italy}\address{INFN, Laboratori Nazionali di Frascati, Via E.
          Fermi 40, I-00044 Frascati, Italy}\thanks{Partially
          supported by EU IHP-RTN contract HPRN-CT-2002-00311 (EURIDICE)},
        A.S.B.~Tariq\addressmark[SU]\address{Department of Physics,
          Rajshahi University, Rajshahi 6205,
          Bangladesh}\thanks{Supported by a Commonwealth Scholarship}}
       
\begin{document}

\begin{abstract}
We report calculations of $\bk$ using two flavours of dynamical
clover-improved Wilson lattice fermions and look for dependence on the
dynamical quark mass at fixed lattice spacing. We see some evidence
for dynamical quark effects. In particular $\bk$ decreases as the sea
quark masses are reduced towards the up/down quark mass. Our meson
masses are quite heavy and a firm prediction of $\bk$ is a task for
future simulations.
\end{abstract}

\let\oldrightmark\rightmark
\def\rightmark{{\rm SHEP--0427, RM3--TH/04--21}}

\maketitle

\section{INTRODUCTION}

$\bk$ determines the value of the $\Delta S=2$ neutral kaon mixing
matrix element according to
\begin{equation}
\label{eq:bkdef}
\langle\overline{K^0}|Q^{\Delta S=2}(\mu)|K^0\rangle
 = \frac{8}{3}\bk(\mu)f_K^2m_K^2,
\end{equation}
with $\mu$ indicating the scale dependence of the operator $Q^{\Delta
S=2}(\mu) = \sbar{\g_\mu(1-\g_5)}d \;\sbar\g^\mu(1-\g_5)d$.

The quenched lattice value of $\bk$ has more or less settled down
(see~\cite{Gupta:2003hu} for a recent review) and the 1997 staggered
fermion result~\cite{Aoki:1998nr} $\bk(\msbar,2\gev) = 0.63(4)$,
corresponding to a renormalisation group invariant (RGI) value $\bkh =
0.87(6)$, remains the benchmark. The quoted error does not include
quenching effects which have been estimated to be as high as
$\pm15\%$~\cite{Sharpe:1997ih} and remain the primary systematic
effect to be addressed. Here we report on a
calculation~\cite{bknf2-2004} using two degenerate flavours of
dynamical (clover-improved) Wilson fermions, aiming to look for
sea-quark dependence in $\bk$.

\section{SETUP AND SIMULATION}
\label{sec:setup}

The operator of interest in the continuum is
\begin{equation}
Q_1(\mu)=\sbar\g_\mu d\;\sbar\g^\mu d+\sbar\g_\mu\g_5d\;\sbar\g^\mu\g_5d,
\end{equation}
which is the parity conserving part of $Q^{\Delta S=2}(\mu)$ in
eq.~(\ref{eq:bkdef}). For Wilson fermions, owing to the explicit
breaking of chiral symmetry, this operator mixes with other
four-fermion operators. Including the overall multiplicative
renormalisation, the subtraction of the unwanted operators may be
expressed as
\begin{equation}
\label{eq:subtr}
Q_1(\mu) = Z (\mu, g_0^2) \,\Big( Q_1^{\mathrm{latt}}
  + \sum_{i\neq 1} \Delta_i(g_0^2)  Q_i^{\mathrm{latt}} \Big).
\end{equation}
The coefficients $Z$ and $\Delta_i$ have been determined
perturbatively to ${\cal O}(\as)$ for $\msbar$-NDR in
\cite{Gupta:1997yt,Capitani:1998nj}. The $\Delta S=2$ lattice operator
can now be matched to the continuum one, having the same chiral
properties up to ${\cal O}(a^0)$. There can be remnant operator
mixing, beginning at ${\cal O}(\as^2 a^0)$, because the $\Delta_i$
coefficients are not known exactly. Even though we use an
improved-clover action, the four-fermion operators are unimproved and
${\cal O}(a)$ artefacts can also be present.

By calculating matrix elements with non-zero momentum kaons we
introduce another degree of freedom, allowing us partially to remove
the lattice artefacts. The chiral behaviour of the matrix element can
be written as~\cite{Sharpe:1997ih}
\begin{eqnarray}
\label{eq:bpchi}
\lefteqn{\langle \bar P^0, \vec p | Q_1(\mu) | P^0, \vec q \rangle
 = \alpha ' +\beta ' m^2_P + \,\delta ' m_P^4 +\mbox{}} \nonumber\\
 & & p{\cdot}q \,\big[\gamma+\gamma ' +(\epsilon+\epsilon ')m_P^2 +
     (\xi+\xi ')p{\cdot}q \big] + \cdots
\end{eqnarray}
The parameters $\g'$, $\epsilon'$ and $\xi'$ are corrections to the
corresponding physical contributions. In contrast, $\alpha '$,
$\beta'$ and $\delta'$ are artefacts of chiral symmetry breaking, in
this case ${\cal O}(a,\as^2 a^0)$ with $\as^2$ small, which have to be
subtracted. In particular the $\alpha'$ term makes $\bk$ diverge in
the chiral limit. Following~\cite{Sharpe:1997ih}, we ignore chiral
logarithmic terms in eq.~(\ref{eq:bpchi}), given the errors and large
masses used here. For our calculation we neglect higher order terms
and use the following expression for the matrix elements:
\begin{equation}
\label{eq:bpwilfit}
\langle \bar P^0, \vec p| Q_1(\mu)| P^0, \vec q\rangle
 = \alpha' +\beta' m^2_P + (\gamma+\gamma')p{\cdot}q .
\end{equation}
 
We calculate $\bk$ using Clover-improved Wilson
fermions~\cite{Luscher:1997ug} on the UKQCD set of unquenched
configurations~\cite{Irving:2000hs,Allton:2001sk} listed in
table~\ref{tab:configs}. We use three different sea quark masses in
the region $m_P/m_V\ge 0.7$ on a volume of $16^3\times 32$ ($m_{P}
L\ge 7$) with a nearly constant lattice spacing determined from the
Sommer scale, $r_0$.
\begin{table}
\setlength{\tabcolsep}{5.5pt}
\caption{The configurations used~\cite{Irving:2000hs,Allton:2001sk}.
  Lattice spacings are fixed from $r_0$. $m_P/m_V$ is quoted for
  $\ksea=\kval$.}
\label{tab:configs}
\begin{tabular}{@{}cccccc@{}}
\hline
$\beta$& $c_\mathrm{SW}$ & $\kappa_\mathrm{sea}$ &
$a/\mathrm{fm}$ & $m_P/m_V$ & \#cfgs \\ 
\hline
5.20 & 2.0171 & 0.1350 & 0.103(2) & 0.70(1) & 100  \\ 
5.26 & 1.9497 & 0.1345 & 0.104(1) & 0.78(1) & 100  \\ 
5.29 & 1.9192 & 0.1340 & 0.102(2) & 0.83(1) & \phantom{0}80   \\ 
\hline
\end{tabular}
\end{table}
Propagators and correlators are calculated using the
FermiQCD~\cite{DiPierro:2001yu,DiPierro:2000ve} code. Five valence
quark propagators at $\kappa =0.1356$, $0.1350$, $0.1345$, $0.1340$
and $0.1335$ are generated for each sea quark. We follow the standard
procedure~\cite{Gavela:1988bd} to evaluate the matrix element from 3-
and 2-point correlation functions.
\begin{eqnarray}
\label{eq:corr}
\lefteqn{{\cal C}^{(3)}(t_x,t_y;p_x,p_y;\mu) = \nonumber}\\ 
 & &  \sum_{\vec x,\vec y} \langle P_5(\vec x,t_x) Q(\vec 0;\mu)
   P_5(\vec y,t_y)\rangle e^{i \vec p_x \vec x}\,e^{i \vec p_y \vec y}, \\
\lefteqn{{\cal C}^{(2)}_{J_iJ_j}(t;p_x) = 
 \sum_{\vec x}\langle\,J_i(\vec x,t)J_j^\dagger(0,0)\rangle
 e^{i \vec p_x\vec x}.}
\end{eqnarray}
We use local pseudoscalar sources $J_i$ for the kaons (smearing with
the method of~\cite{Boyle:1999gx} did not improve the signal
significantly). In the 3-point functions, $Q$ is at the origin and
$t_y$ is fixed at $10$ while $t_x$ is varied over the full temporal
range. We have checked neighbouring values of $t_y$ but observe no
dependence, implying that the ground state is reasonably well
isolated. The momentum configurations $\{\vec p_x,\vec p_y\}$ are
chosen from the pairs $\{(0,0,0),(0,0,0)\},\{(0,0,0),(1,0,0)\}$ and
$\{(1,0,0),(0,0,0)\}$ where an average over equivalent configurations
is implied. The matching coefficients are evaluated using the boosted
coupling, $g_0^2=6\langle P\rangle/\beta$, where the average plaquette
values are $\langle P\rangle \in \{0.5336,0.5399,0.5424\}$ on our
ensembles.

\section{ANALYSIS AND DISCUSSION}
\label{sec:disc}

The matrix element is extracted using the ratios:
\begin{eqnarray}
\label{eq:ratio}
\lefteqn{R_3(\vp) = \frac{{\cal C}^{(3)}(t_x,t_y;p_x,p_y;\mu)}
            {Z^2_A{\cal C}^{(2)}_{PP}(t_x;p_x)
            {\cal C}^{(2)}_{PP}(t_y;p_y)}}\\
\lefteqn{X(0) = \frac{8}{3}\left|
      \frac{{\cal C}^{(2)}_{A_0P}(t_x)}{{\cal C}^{(2)}_{PP}(t_x)}
      \right|^2,
\quad
X(\vp) = X(0)\,\frac{p_x{\cdot}p_y}{m_P^2}}
\end{eqnarray}
where $Z_A$ is the axial current renormalisation. Our Method~I is to
follow~\cite{Becirevic:2002mm,Crisafulli:1996ad} and fit
the equation
\begin{equation}
\label{eq:romefit}
R_3(\vp) = \widetilde\alpha '+\widetilde\beta 'X(0)
      +(\,\widetilde\g + \widetilde\g'\,) X(\vp),
\end{equation}
to find $\bk$ from $\widetilde\g$, neglecting $\widetilde\g'$. This
gives an estimate of the leading term in an expansion of $\bk$ for
each set of different valence quarks with a given sea quark mass, with
the kaons not necessarily being at the physical kaon mass. Using a
linear fit versus the unitary pseudoscalar masses,
$(am_P)^2|_{\ksea=\kval}$, to go to the up/down limit gives
\begin{equation}
\label{eq:bkmethodI}
\bk(\msbar,2\gev) = 0.49(13),
\end{equation}
corresponding to the RGI value $\bkh = 0.69(18)$. In this method
the pseudoscalar masses are for valence quarks in the simulated region
($m_q \approx 2 m_s$) and the $\bk$ estimate is one where the sea
quarks are realistically light but the valence quarks are heavier than
the physical strange quark.

\begin{figure}
\includegraphics[width=\hsize,bb=51 50 517 431]{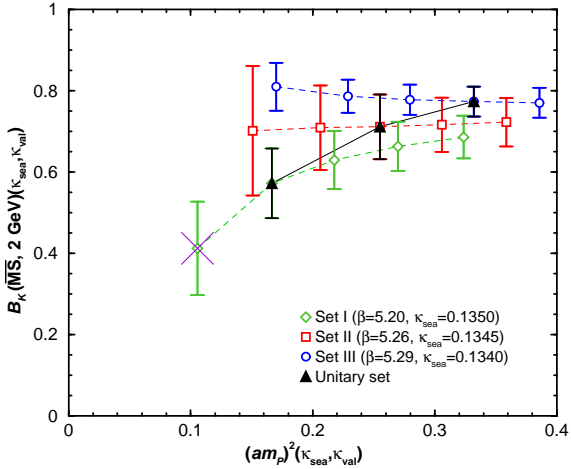}
\caption{$\bk(\msbar, 2\gev)$ for each $(\ksea,\kval)$ as a function
  of the squared pseudoscalar mass. Dashed lines are for guidance to
  distinguish sets with different sea quarks. The filled points joined
  by a solid line have $\ksea=\kval$. The lightest point (marked by a
  large cross) is excluded from the analysis.}
\label{fig:bkkskv}
\end{figure}

In Method~II we follow~\cite{Gupta:1997yt,Gupta:1993bd} and estimate
$\bk$ for each $(\ksea,\kval)$ combination:
\begin{equation}
\label{eq:bkkskv}
\left.\frac{R_3(\vp)-R_3(0)}{X(\vp)-X(0)}\right|_{(\ksea,\kval)}
  = \bk(\mu,\ksea,\kval).
\end{equation}
The resulting values are illustrated in fig.~\ref{fig:bkkskv}.
Taking the unitary points (as done in~\cite{Izubuchi:2003rp}), with
$\ksea=\kval$, and extrapolating to the physical kaon mass results in,
\begin{equation}
\label{eq:bkmethodII}
 \bk(\msbar,2\gev) = 0.48(13).
\end{equation}
Here we have $m_P=m_K^{\mathrm{phys}}$, but the sea and valence quarks
are degenerate and hence the sea content is not as light as the
up/down quarks. Moreover, the kaon comprises two quarks around
$m_s/2$.

Although we recognise the presence of lattice artefacts, it seems that
dynamical quark effects can lower the value of $\bk$. Taking this with
the observation~\cite{Soni:1996qq} that $N_f=2$ results are always
below those for $N_f=0$, a statement also valid for subsequent works,
we see that when one has two still-heavy sea quarks, $\bk$ starts to
decrease but is consistent with the quenched value. When the sea
quarks are taken towards the massless limit, $\bk$ becomes distinctly
lower. A similar effect is seen in the domain-wall results
of~\cite{Izubuchi:2003rp,RBC-dawson-lat04}. In contrast, preliminary
$2{+}1$ flavour results with improved staggered
quarks~\cite{Gamiz:2004} show some sign of a decrease in $\bk$ with
decreasing sea-quark mass, but the overall result is consistent with
the quenched value. It is intriguing to note that recent Unitarity
Triangle fits, which determine $\bkh$ after imposing the remaining
constraints, also find a value below the usual quenched lattice result
(for example, $\bkh = 0.65(10)$ in~\cite{Bona:2004sj}). However, our
primary result is a decrease in $\bk$ as the sea-quark mass decreases,
rather than a firm value for use in phenomenology.


\let\rightmark\oldrightmark

\noindent
Acknowledgment: we thank Ivan Wolton, Ian Hardy, Oz Parchment and the
\emph{Iridis} team at the University of Southampton for computing
support.

\bibliographystyle{espcrc}
\bibliography{bk}

\end{document}